\documentclass[letterpaper, journal]{IEEEtran}

\usepackage{amsthm,amsmath,amssymb,color,graphicx,caption,subcaption,tikz,url,latexsym,xfrac,bm} \usetikzlibrary{decorations.markings,shapes.geometric}
\usepackage{cite}
\usepackage{multirow}
\usepackage[draft,pdfborder={0 0 0}]{hyperref}

\usepackage{colortbl}
\usepackage{pgfplots}
\usepgfplotslibrary{units}
\usetikzlibrary{spy,backgrounds,positioning,patterns}

\usepackage{nicefrac}
\usepackage{glossaries} 
\usepackage[linesnumbered, ruled, vlined, noend]{algorithm2e}

\definecolor{darkred}{rgb}{0.75, 0, 0}
\definecolor{darkgreen}{rgb}{0, 0.45, 0}
\definecolor{bluegreen}{rgb}{0, 0.84, 0.84}
\definecolor{matlab1_blue}{rgb}{0, 0.447, 0.741} 
\definecolor{matlab2_red}{rgb}{0.850, 0.325, 0.098} 
\definecolor{matlab3_yellow}{rgb}{0.929, 0.694, 0.125} 
\definecolor{matlab4_violet}{rgb}{0.494, 0.184, 0.556} 
\definecolor{matlab5_green}{rgb}{0.466, 0.674, 0.188} 
\definecolor{matlab6_lightblue}{rgb}{0.301, 0.745, 0.933} 
\definecolor{matlab7}{rgb}{0.635, 0.078, 0.184} 
\newcommand{\ubf}{\mathbf{u}}
\newcommand{\xbf}{\mathbf{x}}
\newcommand{\ybf}{\mathbf{y}}
\newcommand{\mbf}{\mathbf{m}}

\newcommand{\Acal}{\mathcal{A}}

\newcommand{\Ical}{\mathcal{I}} 
\newcommand{\Fcal}{\mathcal{F}} 

\newcommand{\BLTA}{\mathsf{BLTA}}

\newcommand{\Pbb}{\mathbb{P}}
\newcommand{\Ebb}{\mathbb{E}}
\newcommand{\given}{\,\vert\,}

\DeclareMathOperator{\adec}{aSC}
\DeclareMathOperator{\SC}{SC}

\newcolumntype{C}[1]{>{\centering\let\newline\\\arraybackslash\hspace{0pt}}m{#1}}

\newcommand{\Mfo}{\overline{M}_{\mathrm{fo}}}

\newif\ifshowchanges
\showchangesfalse  

\newcommand{\change}[1]{%
  \ifshowchanges\textcolor{blue}{#1}\else#1\fi%
}

\begin{document}
\bstctlcite{BSTcontrol}    

\newcommand{\mypoint}[2]{\tikz[remember picture]{\node[inner sep=0, anchor=base](#1){$#2$};}}

\newacronym{ae}{AE}{automorphism ensemble}
\newacronym{aed}{AED}{automorphism ensemble decoder}
\newacronym{awgn}{AWGN}{additive white gaussian noise}
\newacronym{ber}{BER}{bit-error-rate}
\newacronym{bler}{BLER}{block-error rate}
\newacronym{bpsk}{BPSK}{binary phase-shift keying}
\newacronym{blta}{BLTA}{block-lower-triangular affine}
\newacronym{dae}{DAE}{dynamic automorphism ensemble}
\newacronym{ga}{GA}{general affine}
\newacronym{llr}{LLR}{log-likelihood ratio}
\newacronym{lsm}{LSM}{least-squares metric}
\newacronym{ml}{ML}{maximum likelihood}
\newacronym{pdae}{PDAE}{partial dynamic automorphism ensemble}
\newacronym{pm}{PM}{path metric}
\newacronym{rm}{RM}{Reed-Muller}
\newacronym{sc}{SC}{successive cancellation}
\newacronym{snr}{SNR}{signal-to-noise ratio}
\newacronym{upo}{UPO}{universal partial order}

\title{Sequential Automorphism Ensemble Decoding\\with Early Stopping}
\author{Charles Pillet, Pascal Giard, Bassant Selim and François Leduc-Primeau
\thanks{Charles Pillet and Bassant Selim (charles.pillet@lacime.etsmtl.ca, bassant.selim@etsmtl.ca) are with Synchromedia, Depart.\@ of Systems Engineering,  École de technologie supérieure, Montréal, QC, Canada.
Pascal Giard (pascal.giard@etsmtl.ca) is with LaCIME, Depart.\@ of Electrical Engineering, École de technologie supérieure, Montréal, QC, Canada. 
François Leduc-Primeau (francois.leduc-primeau@polymtl.ca) is with the Depart.\@ of Electrical Engineering, Polytechnique Montréal, Montréal, QC, Canada.}}

\maketitle
 
 \begin{abstract}
In this paper, a low-complexity approach for the \gls{aed} using \gls{sc} as constituent decoders is proposed.
The approach sequentially activates sub-decoders and terminates the decoding process based on pre-optimized parameters, derived from the strong correlation observed between the decoding outcome and the \gls{sc} path metric.
An algorithm is proposed to find a list of early termination thresholds that minimize average decoding complexity subject to a \gls{bler} constraint.
For various code parameters and a \gls{bler} below \(10^{-3}\), simulation results show that average decoding complexity is reduced by a factor of at least $6\times$, and up to $22\times$,
compared to the original \gls{aed} complexity, with a negligible degradation in \gls{bler}.
\end{abstract}
\glsresetall
	\IEEEpeerreviewmaketitle	
	\section{Introduction}
    \gls{rm} codes \cite{Reed}\cite{Muller}, one of the oldest channel codes, are known for their good \gls{ml} performance in the finite-length regime.
    \gls{rm} codes have regained attention due to their strong connection with polar codes, seen as a breakthrough since their publication \cite{ArikanPolarCodes}.
    Their construction ensures that they asymptotically achieve the capacity of a discrete memoryless channel under the low-complexity \gls{sc} decoding algorithm.
    
    Recently, a new family of decoders has emerged for codes that share properties with polar and \gls{rm} codes \cite{PSMC}.
    This family uses an ensemble of independent decoders to return the most likely codeword estimate  based on an \gls{ml} criterion.
    \change{These decoders are automorphism decoders \cite{AEjournal}, i.e., they decode} a permuted version of the channel realization, where the permutations are automorphisms of the code. 
    This decoder is labeled as the \gls{ae} decoding algorithm \cite{geiselhart2020automorphism}.

    \gls{ae} decoding was first proposed for \gls{rm} codes \cite{geiselhart2020automorphism}.
    \gls{rm} codes exhibit key properties, such as a large minimum distance and large affine automorphism group \change{$\mathcal{A}$}, that ensure good decoding performance under \gls{ae} decoding.
    For example, under \gls{ae} decoding, the \gls{rm} code of size $N=128$ and dimension $K=64$ was shown to achieve \gls{ml} performance with $32$ \gls{sc} decoders \cite{geiselhart2020automorphism}.
    Polar codes following the universal partial order \cite{partial_order} also have a known affine automorphism group, isomorphic to the \gls{blta} group \cite{geiselhart2021automorphismPC}.

    \change{The absorption group, a subgroup of $\mathcal{A}$, is of particular interest. Applying an automorphism from this subgroup to an SC decoder generates the same decoding outcome as directly using SC \cite{geiselhart2020automorphism, SC_invariant}, defining an equivalence relation on $\mathcal{A}$. Consequently, the automorphism group $\mathcal{A}$ can be split into equivalence classes and the set of SC-variant automorphisms can be reduced by retaining a single representative per equivalence class \cite{AE_v2,AEjournal}.}
    \change{Nevertheless, the number of equivalence classes} is usually higher than the ensemble size used in \gls{ae}.

    Various works have investigated polar code construction with a desired affine automorphism group. It was shown that such codes are available in almost all code dimensions, and exhibit a large enough automorphism group to improve their error-correction performance under \gls{ae} with \gls{sc} as the constituent decoder\cite{geiselhart2021automorphismPC,AE_v1,PSC,RateCompatibleAE,pre_defined_auto}.

    The \gls{ae} decoder can be implemented in hardware by instantiating the constituent decoders in parallel to reduce latency \cite{AESC_HW}, or by re-using a single instance sequentially to reduce area usage~\cite{rubenackeSerialAutoEnsemble}.
    The main drawback of \gls{ae} decoding is of course that, it increases decoding complexity compared to a single SC decoder. 
    Early-termination in \gls{ae} decoding was proposed in \cite{quasi_optimal_PCA,SEA_AED}, both based on path metrics.
    In \cite{quasi_optimal_PCA}, all sub-decoders are executed in parallel while some of them can be early terminated by comparing the partial decoding results among themselves. 
    Results are only presented for ensembles of size 32 or larger and only for \gls{rm} codes.
    In \cite{SEA_AED}, the authors simplified an early-termination approach proposed in \cite{PermDecRussian} by using the path metrics rather than the whole codeword.
    
    In this letter, we introduce another early-termination scheme based on the correlation of the path metric value and the decoding outcome.
    The proposed stopping condition is evaluated by comparing the SC path metric with precomputed parameters denoted as $\bm{\sigma}$. 
    These parameters are tailored to minimize the complexity while satisfying a desired \gls{bler} constraint.
    The resulting energy-efficient decoder, labeled as the \gls{dae} decoder, greatly reduces the average complexity at the cost of variable execution time. 
    Compared to \cite{quasi_optimal_PCA,SEA_AED,PermDecRussian}, the proposed \gls{dae} method achieves better complexity reductions, while explicitly guaranteeing the \gls{bler}.
    
    \section{System Model}\label{sec:prelim}
    \subsection{Polar and Reed-Muller codes, and Successive Cancellation}
    Polar  and \gls{rm} codes of size $N=2^n$ are two classes of binary linear codes that can be described with the matrix $\mathbf{G}_N$ defined as $\mathbf{G}_N=\left[\begin{smallmatrix}1 &0\\ 1 &1\end{smallmatrix}\right]^{\otimes n}$, where $(\cdot)^{\otimes n}$ denotes the $n^{\text{th}}$ Kronecker power \cite{ArikanPolarCodes}.
    Both codes are defined by the information set $\mathcal{I}$, containing the $K$ indices storing the $K$ bits of information in an input vector $\ubf$.
    The remaining indices are stored in the frozen set $\mathcal{F}$.
    The encoding is performed as 
    \begin{align}
        \xbf=\ubf\mathbf{G}_N,
    \end{align}
    where $\xbf\in\mathbb{F}_2^N$ is the codeword, and $\ubf\in\mathbb{F}_2^N$ is the input vector, generated by assigning \change{$\ubf_\Fcal=\mathbf{0}$ and $\ubf_\Ical=\mbf$}, where $\mbf\in\mathbb{F}_2^K$ is the $K$-bit vector of information to transmit. 
    The difference between polar and \gls{rm} codes resides in the choice of the indices in the information set $\Ical$.
    For a given code dimension $K$ and a channel condition, a polar code is constructed by picking the $K$ most reliable positions \cite{AWGN_construction}.
    A \gls{rm} code with parameters $r$ and $n$ constructs the information set $\Ical$ as all indices $k$ whose binary representation have at least $n-r$ ones.
    Hence, \gls{rm} codes of size $N=2^n$ are defined only for a limited set of code dimensions $K$.

    \Gls{sc} is the original soft-input/hard-output decoding algorithm of polar codes  \cite{ArikanPolarCodes}.
    \change{\Gls{sc} can also be used to decode \gls{rm} codes \cite{SC_RM}.}
    Its schedule decodes one bit at a time, by propagating \glspl{llr} using $f$ or $g$ functions.
    The estimated bit $\hat{u}_k$ is either a hard decision based on the propagated \gls{llr} on indices $k\in\Ical$ or $\hat{u}_k=0$ if $k\in\Fcal$.
    Choosing the most reliable positions for $\Ical$ ensures that polar codes are capacity achieving under \gls{sc} as $N\rightarrow\infty$.
    Due to its known application in \gls{ae} \cite{geiselhart2021automorphismPC,SC_invariant,AE_v2}, \gls{sc} will be used in this letter.
         
    \subsection{Automorphism Ensemble Decoding}
    Based on a channel realization $\ybf$ of the transmitted codeword $\xbf$, the corresponding automorphism decoder $\adec$ of an \gls{sc} decoder \cite{AEjournal}, is given by
    \begin{equation}
    \label{eq:adec}
    \adec(\ybf,\pi) = \pi^{-1}\left( \SC(\pi(\ybf)) \right),
    \end{equation}
    where $\pi$ is an automorphism.
    The  decoder $\adec$ returns an estimate $\hat{\xbf}$ of the codeword $\xbf$. 
    Given two automorphisms $\pi_1$, $\pi_2$ in the same equivalence class, both $\adec$ return the same $\hat{\xbf}$ for any $\ybf$.
    The number of equivalence classes is derived from the size of \change{the affine automorphism group} $\Acal$ and the size of the absorption group, studied in \cite{geiselhart2020automorphism,SC_invariant,AEjournal}.

    The \gls{ae} decoding algorithm, labeled as AE-$M$-SC, is an ensemble of $M$ independent $\adec$ decoders based on $M$ automorphisms $\pi_1,\pi_2,\dots,\pi_M$ taken from $M$ distinct equivalence classes.
    The $M$ estimated codewords, stored in a set $\mathcal{X}=\{\hat{\xbf}_1,\dots,\hat{\xbf}_M\}$ are compared with an \gls{ml} criterion for the \gls{awgn} channel \cite{geiselhart2020automorphism}.

    A reliability metric for the SC decoder called the \gls{pm} was proposed in
    \cite{AESC_HW}.
    We denote $m_i$ the \gls{pm} of the $i^{\text{th}}$ sub-decoder.
    The \gls{pm} is first initialized to 0, and then updated during SC decoding.
    When reaching the leaf node associated with $\hat{u}_k$ in the \gls{sc} decoding tree, $m_i$ is updated as
    \begin{align}
    m_i\gets 
        \begin{cases}
        m_i,& \text{if}\,\, k\in\Ical,\\
        m_i,& \text{if}\,\,k\in\Fcal\wedge \text{HD}(L_k)= 0, \\
        m_i+ \left|L_k\right|, \,& \text{if}\,\,k\in\Fcal\wedge \text{HD}(L_k)\neq 0.
        \label{eq:pm_ae}
        \end{cases}
    \end{align}
    Hence, the metric is penalized if the hard decision $\text{HD}(L_k)$ on the \gls{llr} of the $k^{\text{th}}$ leaf disagrees with the known value $u_k=0$ for indices $k\in\Fcal$. 
    Since the decoder does not know \emph{a priori} the value $u_k$ for indices $k\in\Ical$, the metric is left unchanged.

    The output $\hat{\xbf}^{(M)}\in\mathcal{X}$ of the AE-$M$-SC decoder is selected based on the \glspl{pm}, as 
    \begin{equation}
    \hat{\xbf}^{(M)}=\hat{\xbf}_k,\, k=\underset{j=1,\dots,M}{\arg\min}\; m_j \,.\label{eq:decision}
    \end{equation}
    
    \section{Low-Complexity Sequential \gls{ae} Decoding}\label{sec:AED}
    As evident from \eqref{eq:decision}, the final output returned by the \gls{ae} decoder only depends on one of the constituent decoders. Therefore, its decoding complexity could be reduced if one could reliably predict which automorphism will lead to the correct decoding outcome for a particular received vector. However, making such a prediction with computational complexity smaller than the ensemble of SC decoders is potentially difficult.
    In this letter, we consider instead that the \change{sequence} of automorphisms is fixed, and attempt to predict the number of sub-decoders that should be used for each received vector.

    \subsection{\change{Fixed-Sequence Oracle Decoder}}\label{subsec:oracle_DAE}
    To support the approach and provide \change{a} baseline, we consider \change{an} oracle decoder that \change{has prior} information about the decoding outcomes, which \change{it} can exploit to reduce computational complexity, defined as the expected number of sub-decoders used \change{over the channel realizations}, while being functionally equivalent to the AE-$M$-SC decoder, that is, always returning $\hat{\xbf}^{(M)}$ as in \eqref{eq:decision}.

    \change{The} \change{\emph{fixed-sequence oracle}} (FO)  executes the sub-decoders in a fixed \change{sequence} while being able to perfectly detect whether a decoding outcome is equal to $\hat{\xbf}^{(M)}$.
    \change{To express the complexity of FO, let us denote by $n_\textsc{ml}$ the number of sub-decoders that output $\hat{\bf x}^{(M)}$.}
    \change{Since applying an automorphism does not modify the code nor the noise realization}, all sub-decoders have the same probability of returning $\hat{\xbf}^{(M)}$, and the complexity of FO can be expressed in terms of $n_\textsc{ml}$ as
    \begin{equation}\label{eq:FOcomplexity}
        \Mfo = 
        \change{
        \sum_{\nu=1}^M\frac{M+1}{\nu+1} \Pbb(n_\textsc{ml}=\nu)}\,.
    \end{equation}
    \change{The expression above is obtained by considering that given $n_\textsc{ml}$, the complexity corresponds to the expected number of trials until the first success when drawing without replacement from a set of size $M$ containing $n_\textsc{ml}$ successes.} 
    
    \begin{figure}
    \captionsetup[subfigure]{skip=1pt}
        \centering
        \begin{subfigure}{0.99\columnwidth}
            \resizebox{0.9\columnwidth}{!}{\begin{tikzpicture}
\begin{axis}[
    ybar, 
    width=\columnwidth,
    height=0.425\columnwidth,
    xlabel={$n_\textsc{ml}$}, 
    ylabel={$\mathbb{P}(n_\textsc{ml})$}, 
    symbolic x coords={1, 2, 3, 4, 5, 6, 7, 8}, 
    bar width=5, 
    ymin=0,
    ymax=1.01,
    xtick=data, 
    legend style={at={(0.01,0.99)},anchor=north west},
    legend style={legend columns=1, font=\small, row sep=-1.2mm},
    legend style={fill=white, fill opacity=1, draw opacity=1,text opacity=1}, 
    legend cell align={left}, 
    mark size=1.6pt, mark options=solid,
    grid=both,
]

\addlegendimage{empty legend}
\addlegendentry{$K\,-\, \Mfo$}

\addplot[fill=matlab1_blue] coordinates{(1,0.002965) (2,0.006135) (3,0.01235) (4,0.02468) (5,0.04904) (6,0.09458) (7,0.1988) (8,0.6115)  };
\addlegendentry{$29\,-\,1.125$}
\addplot[pattern=north east lines,
    pattern color=matlab2_red] coordinates {(1,0.003983) (2,0.007314) (3,0.01318) (4,0.0236) (5,0.046) (6,0.08015) (7,0.1592) (8,0.6665) };
\addlegendentry{$64\,-\,1.129$}
\addplot[fill=matlab3_yellow!60] coordinates {(1,0.001806) (2,0.002756) (3,0.004889) (4,0.008939) (5,0.01556) (6,0.02699) (7,0.05314) (8,0.8859)};
\addlegendentry{$99\,-\,1.047$}

\end{axis}
\end{tikzpicture}}
            \caption{RM codes of size $N=128$.}
            \label{fig:redundancy_bar_001}
        \end{subfigure}
        
        \vspace{5pt}
        
        \begin{subfigure}{0.99\columnwidth}
            \resizebox{0.9\columnwidth}{!}{\begin{tikzpicture}
\begin{axis}[
    ybar, 
    width=\columnwidth,
    height=0.425\columnwidth,
    xlabel={$n_\textsc{ml}$}, 
    ylabel={$\mathbb{P}(n_\textsc{ml})$}, 
    symbolic x coords={1, 2, 3, 4, 5, 6, 7, 8}, 
    bar width=5, 
    ymin=0,
    ymax=1.01,
    xtick=data, 
    legend style={at={(0.01,0.99)},anchor=north west},
    legend style={legend columns=1, font=\small, row sep=-1.2mm},
    legend style={fill=white, fill opacity=1, draw opacity=1,text opacity=1}, 
    legend cell align={left}, 
    mark size=1.6pt, mark options=solid,
    grid=both,
]

\addlegendimage{empty legend}
\addlegendentry{$K\,-\, \Mfo$}

\addplot[fill=matlab1_blue] coordinates{(1,0.0006474) (2,0.001538) (3,0.003009) (4,0.006294) (5,0.0134) (6,0.03222) (7,0.05723) (8,0.8857)};
\addlegendentry{$23\,-\,1.037$}
\addplot[pattern=north east lines,
    pattern color=matlab2_red] coordinates {(1,0.002093) (2,0.003706) (3,0.006524) (4,0.01202) (5,0.02241) (6,0.04349) (7,0.1023) (8,0.8074)};
\addlegendentry{$60\,-\,1.069$}
\addplot[fill=matlab3_yellow!60] coordinates {(1,0.001655) (2,0.002509) (3,0.003928) (4,0.006282) (5,0.01003) (6,0.0159) (7,0.02766) (8,0.932)};
\addlegendentry{$98\,-\,1.034$}

\end{axis}
\end{tikzpicture}}
            \caption{Polar codes of size $N=128$.}
            \label{fig:redundancy_bar_001_POLAR}
        \end{subfigure}        
        \caption{$\Pbb(n_\textsc{ml})$ for $M=8$ and BLER of $10^{-3}$.}\label{fig:redundancy_bar}
    \end{figure}
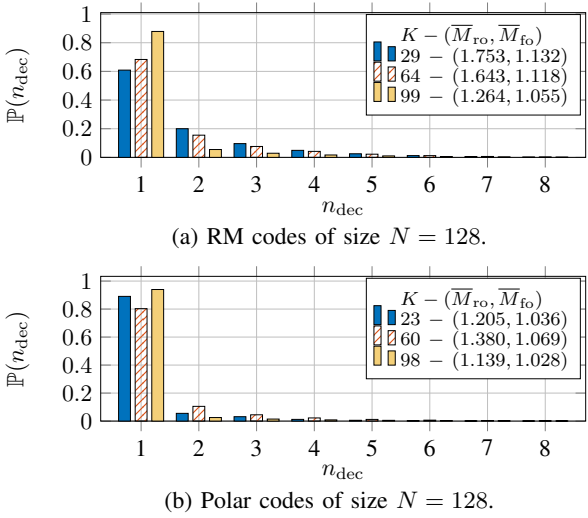

    We present empirical results for a few relevant codes showing the distribution of \change{$n_\textsc{ml}$} and the resulting oracle complexity.
    \autoref{fig:redundancy_bar_001} considers 3 \gls{rm} codes  with code dimension $K=\{29,64,99\}$ and \autoref{fig:redundancy_bar_001_POLAR} considers 3 polar codes  with code dimension $K=\{23,60,98\}$ and $\Acal=\BLTA(3,4)$. All 6 codes have length $N=128$ and transmission takes place on an \gls{awgn} channel with noise adjusted for a \gls{bler} of $10^{-3}$.
    The $M=8$ automorphisms have been picked randomly while ensuring they are in $M$ distinct equivalence classes.

    For all 6 codes, \change{$n_\textsc{ml}=M$} has high probability, showing the potential for complexity reduction.
    For \gls{rm} codes, the probability of \change{$n_\textsc{ml}=M$} increases with the code rate.
    We also notice that the polar codes exhibit higher redundancy than the \gls{rm} codes, which can be explained by their smaller affine automorphism group, resulting in less effective scrambling.
    
\subsection{Correlation Between \change{Decision Metrics} and Redundancy}
In order to set the stage for the proposed algorithm, we study the correlation between the decoding outcome of an $\adec$ decoder and two common decision metrics for \gls{ae} decoders, namely the \gls{lsm} used in \cite{geiselhart2020automorphism} and the \gls{pm} used in \cite{AESC_HW}.

\begin{figure}
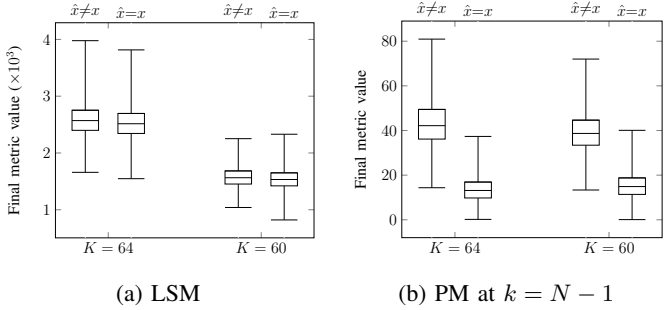

    \centering
    \begin{subfigure}[b]{0.48\columnwidth}
        \centering
        \resizebox{\columnwidth}{!}{\input{figures/subFigBoxplotLSM}}
        \caption{\Gls{lsm}}
        \label{fig:boxplotLSM}
    \end{subfigure}
    \hfill
    \begin{subfigure}[b]{0.48\columnwidth}
        \centering
        \resizebox{\columnwidth}{!}{\input{figures/subFigBoxplotSCM}}
        \caption{\gls{pm} at $k=N-1$}
        \label{fig:boxplotSCM}
    \end{subfigure}
    \caption{Distribution of \gls{lsm} or \gls{pm} metrics, conditioned on the decoding outcome at $\text{BLER}=10^{-3}$, for \gls{rm} and polar codes with $(N,K)=(128,64)$ and $(N,K)=(128,60)$.}
    \label{fig:boxplot}
\end{figure}

For both metrics, a lower metric value indicates a better candidate.
\autoref{fig:boxplot} depicts 8 boxplots, showing median, minimum, maximum, first and third quartiles of the metric values at the end of decoding at $\text{BLER}=10^{-3}$, for a \gls{rm} code with parameters $(128,64)$ and a polar code with parameters $(128,60)$.
The boxplots are conditioned on the decoding outcome.
A striking observation in \autoref{fig:boxplot} is the clear correlation between the decoding outcome and the metric values, evidenced by the well-separated boxplots. 
\change{This correlation is valid for both codes and metric strategies.}
Therefore, \change{both metrics} can serve as a reliable indicator of whether a sequential \gls{ae} decoder requires additional automorphism decoders to find the correct candidate.
\change{Next, the \gls{pm} is used as it is already computed during decoding.}

\subsection{Dynamic Automorphism Ensemble Decoding}
The \gls{dae} decoding algorithm using \gls{sc} as sub-decoder is described next. 
It is a sequential ensemble decoder, i.e., $M$ SC decoders are invoked one at a time. 
The invocation of each decoder is tracked with index $i$, $1\leq i \leq M$.
This ensemble decoder is \emph{dynamic} because, after each decoding attempt, a stopping criterion is evaluated to determine whether to take a decision based on the codeword estimates already collected, or to continue with more SC attempts. When the decoder terminates after $i$ attempts, either because the stopping criterion has been triggered, or because $i=M$, the decoder output $\hat{\xbf}^{(i)}$ is obtained by substituting $M$ with $i$ in \eqref{eq:decision}.

Given the correlation between the \gls{pm} and the decoding outcome, a stopping criterion is defined based on the \gls{pm}. Let
\begin{equation}
    d^{(i)} \triangleq \min_{j=1,\ldots, i} m_j,\label{eq:min_metric}
\end{equation}
be the minimal \gls{pm} after $i$ decoders. The decoder stops after $i$ invocations if the following condition is satisfied:
\begin{equation}
d^{(i)} < \sigma_i \,,\label{eq:early_termination}
\end{equation}
where $\sigma_i$ are parameters to be optimized.  Since the decoder must stop after $M$ SC invocations, we set $\sigma_M=\infty$.
Next, the parameters are combined into a vector $\bm{\sigma}=[\sigma_1,\dots,\sigma_M]$.

\subsection{Optimization of $\bm{\sigma}$} 
The stopping criteria should be chosen to achieve a desired \gls{bler}. Since the possibility of invoking less than $M$ $\adec$ decoders implies some (potentially negligible) degradation in error rate, we consider that the \gls{bler} $\xi_D$ of the dynamic decoder should satisfy
\begin{equation}\label{eq:FER-constraint}
    \xi_D\leq\xi+\epsilon\,, 
\end{equation}
where $\xi$ is the \gls{bler} of the AE-$M$-SC decoder.
The additional \gls{bler} $\xi_D-\xi$ can be decomposed for each decoding step as
$\xi_D-\xi\change{\triangleq}\sum_{i=1}^{M} \xi_i(\bm\sigma)$.
Denoting by $1 \leq U \leq M$ the number of decoders  invoked before stopping, we evaluate $\xi_i(\bm\sigma)$ as
\begin{multline}\label{eq:xi_i}
    \xi_i(\bm\sigma) = \change{\Pbb(U=i) 
        \big(}\Pbb(\hat{\xbf}^{(i)}\neq \xbf \given U=i \land \hat{\xbf}^{(M)}= \xbf)\\
        -\Pbb(\hat{\xbf}^{(i)}=\xbf \given U=i \land \hat{\xbf}^{(M)}\neq \xbf) \change{\big)}\,,
\end{multline}
where the negative term results from the case where the AE-$M$-SC makes a \emph{maximum-likelihood error}, that is $\xbf\in\mathcal{X} \land \hat{\xbf}^{(M)}\neq\xbf$,
and $U$ is related to the minimal \glspl{pm} and the stopping parameters by the following relationship:
\begin{multline}
    U=u \Leftrightarrow d^{(u)}<\sigma_u \,\land\,\\ 
    d^{(1)}\geq\sigma_1 \,\land\, d^{(2)}\geq\sigma_2 \,\land \dots \land\, d^{(u-1)}\geq\sigma_{u-1} \,.
\end{multline}

Denoting the decoding complexity as $\overline{M}(\bm\sigma)=\Ebb[U]$, \change{where the expectation is in terms of channel realizations,} our optimization problem can be expressed as
\begin{equation}\label{eq:optglb}
    \text{minimize } \overline{M}(\bm\sigma) \;
    \text{s.t.} \sum_{i=1}^M \xi_i(\bm{\sigma}) \leq \epsilon \,.
\end{equation}
However, in practice, we will want to solve this problem by relying on Monte-Carlo simulations of the $\adec$ decoders. Therefore let us consider instead a ``Monte-Carlo'' version of the problem where the choice of $\bm\sigma$ is transformed into a decoding error allocation problem at each step.
Consider a dataset $\mathcal{B}$ of noisy received vectors and a subset $\mathcal{B}_S \subseteq \mathcal{B}$ containing the vectors that are successfully decoded by the AE-$M$-SC. The \gls{bler} constraint $\epsilon$ is transformed into an error count constraint $E=\left\lfloor\epsilon\vert\mathcal{B}_S\vert\right\rfloor$, and we denote by $e_i$ the additional errors introduced at sub-decoder $i$ due to early termination. Within a given dataset, a choice of error allocation $\bm{e}=[e_1,\dots,e_M]$ corresponds to a unique $\bm\sigma$, 
thus we can think of $\overline{M}$ as a function of $\bm e$. Therefore problem \eqref{eq:optglb} becomes
\begin{equation}\label{eq:optMC}
    \text{minimize } \overline{M}(\bm e) \;
    \text{s.t.} \sum_{i=1}^M e_i \leq E \,.
\end{equation}
Note that by definition, $\xi_M(\bm\sigma)=0$, and as a result $e_M=0$.

\begin{algorithm}[t]
\resizebox{0.9\linewidth}{!}{%
\begin{minipage}{\linewidth}
\SetAlgoLined
\DontPrintSemicolon
\SetKwInOut{Input}{input}%
\SetKwInOut{Output}{output}%
\SetKw{Return}{return}

\Input{Dataset $\mathcal{B}_S$ for measuring $\overline{M}(\bf e)$, $E$, $T_{\max}$, $\kappa$}
\Output{$\bm e^*$}

$
e^*_i \gets
    \begin{cases}
        \lceil E/(M-1)\rceil & \text{for $1\leq i \leq M-2$,}\\
        E-\sum_{j=1}^{M-2} e^*_j & \text{for $i=M-1$,}\\
        0 & \text{for $i=M$.}
    \end{cases}
$\;

$i_{\min}\gets 1 ;\; j_{\min}\gets 2 ;\; T\gets 0$\;
\While{$i_{\min}\neq j_{\min}$ and $T<T_{\max}$} {
    $\overline{M}_{\min}\gets \infty$\;
    \For{$i \gets 1$ to $M-1$} {
        $\bm e' \gets \bm e^*;\;$ $e'_i \gets e'_i+\kappa$\;
        \If{$\overline{M}(\bm e') < \overline{M}_{\min}$} {
            $\overline{M}_{\min}\gets\overline{M}(\bm e');\;$
            $i_{\min}\gets i$\;
        }
    }
    $e^*_{i_{\min}} \gets e^*_{i_{\min}}+\kappa;\; \overline{M}_{\min}\gets \infty$\;
    \For{$j \gets 1$ to $M-1$} {
        $\bm e' \gets \bm e^*;\;$ $e'_j \gets e'_j-\kappa$\;
        \If{$\overline{M}(\bm e') < \overline{M}_{\min}$} {
            $\overline{M}_{\min}\gets\overline{M}(\bm e');\;$
            $j_{\min}\gets j$\;
        }
    }
    $e^*_{j_{\min}} \gets e^*_{j_{\min}}-\kappa;\; T\gets T+1$\;
}

\Return{$\bm e^*$}

\end{minipage}
}
\caption{Error Allocation Search}
\label{alg:error_alloc}
\end{algorithm}
It is easy to show that achieving the constraint in \eqref{eq:optMC} with equality is sufficient, since $\sigma_i$ is a non-decreasing function of $e_i$, and $\Ebb[U]$ is a non-increasing function of $\sigma_i$. 
We propose to solve \eqref{eq:optMC} \change{offline} using Algorithm~\ref{alg:error_alloc}, which is a form of steepest ascent hill climbing.
We begin by initializing the allocation $\bm e^*$ with errors spread uniformly across the sub-decoders. Then, we perform hill climbing iterations. At each iteration, a search is performed to find the best pair \change{$(e^{*}_i,e^{*}_j)$} that can exchange an allocation of $\kappa\geq 1$ errors such as to minimize the complexity $\overline{M}(\bm e^*)$.
The algorithm stops when the pair \change{$(e^{*}_i,e^{*}_j)$} at a given \change{iteration} $T<T_{max}$ verifies $i=j$.

As an example, applying this algorithm to the $(128,64)$ \gls{rm} code, with $M=8$, $\epsilon=\nicefrac{\xi}{10}$ and $\kappa=5$, stops at iteration 23. At $T=0$, we have $\bm e^*=[71,\dots,71,0]$ for $\overline{M}=1.6358$. At $T=23$, we have $\bm e^*=[116,101,96,71,61,51,1,0]$ and $\overline{M}=1.6017$. Prioritizing the allocation of more errors to the early steps maximizes the early-termination effect on $\overline{M}$.

\subsection{Partial Dynamic Automorphism Ensemble Decoding}
Next, a variant of the \gls{dae} decoder, referred to as the \gls{pdae} decoder, is introduced.
This variant differs from \gls{dae} by adding the ability to abort a sub-decoder, enabling further complexity reduction.
As evident from its definition in \eqref{eq:pm_ae}, the \gls{pm} is non decreasing during an $\adec$ attempt.
Therefore, during the \( i^{\text{th}} \) $\adec$ attempt (with $i\geq2)$, if the \gls{pm} when reaching the $k^{\text{th}}$ leaf already exceeds \( d^{(i-1)} \), the candidate $\hat{\mathbf{x}}_i$ of the corresponding decoding attempt cannot be selected in \eqref{eq:decision}.
Consequently, with \gls{pdae}, this $\adec$ attempt is aborted, reducing overall complexity while achieving the same error-correction performance as \gls{dae}.
When an $\adec$ attempt is aborted at the $k^\text{th}$ leaf, we define the computational complexity of the attempt as the fraction of $f$ and $g$ function evaluations performed during the attempt, compared to a full SC attempt, evaluated as
\begin{equation}
       \mathcal{L}_{f-g}\left(k\right) = \frac{1}{nN}\sum_{s=0}^{n-1} \left( \left \lceil \frac{k+1}{2^s} \right \rceil \cdot 2^s\right).
\end{equation}

    \section{Simulation Results}\label{sec:sim}
    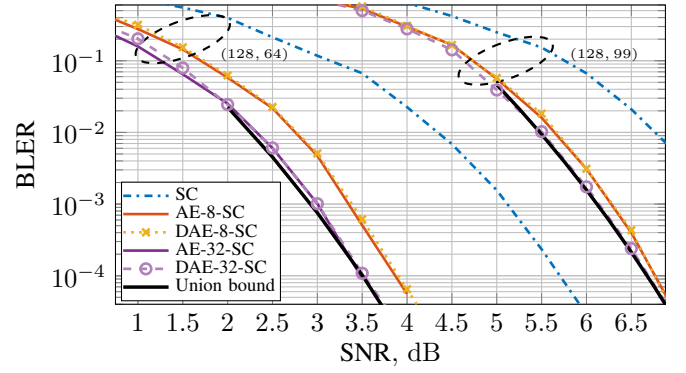
\begin{figure}[t]
        \centering
        \resizebox{0.9\columnwidth}{!}{\begin{tikzpicture}
   \pgfplotsset{
     label style = {font=\fontsize{10pt}{7.2}\selectfont},
     tick label style = {font=\fontsize{10pt}{7.2}\selectfont}
   }

  \begin{semilogyaxis}[%
    width=\columnwidth,
    height=0.625\columnwidth,
    xmin=0.75, xmax=6.88,
    xtick={1,1.5,2,...,7.5},
    xlabel={$\text{SNR},\,\mathrm{dB}$},
    xlabel style={yshift=0.4em},
    ymin=4e-5, ymax=0.6,
    ylabel style={yshift=-0.1em},
    ylabel={BLER},
    yminorticks, xmajorgrids,
    ymajorgrids, yminorgrids,
    legend style={at={(0.01,0.01)},anchor=south west},
    legend style={legend columns=1, font=\scriptsize, row sep=-1.2mm},
    legend style={fill=white, fill opacity=1, draw opacity=1,text opacity=1}, 
    legend style={inner xsep=0pt, inner ysep=0pt}, 
    legend cell align={left}, 
    mark size=1.6pt, mark options=solid,
    ]

    \addplot[color=matlab1_blue, line width=1pt, dashdotted, mark size=2pt]
table[x=snr,y=fer]{figures/BLER/SC_128_64.txt}; 
\addlegendentry{SC}
    \addplot[color=matlab2_red, line width=1pt, mark size=2pt]
table[x=snr,y=fer]{figures/BLER/AE8_128_64.txt};
\addlegendentry{AE-8-SC}
    \addplot[color=matlab5_green,dotted, mark=x, line width=1pt, mark size=2pt]
table[x=snr,y=fer]{figures/BLER/v4/DAE8_128_64_xi10.txt};
\addlegendentry{DAE-8-SC}

    \addplot[color=matlab4_violet, line width=1pt, mark size=2pt]
table[x=snr,y=fer]{figures/BLER/AE32_128_64.txt};
\addlegendentry{AE-32-SC}
    \addplot[color=matlab4_violet!60,dashed, mark=o, line width=1pt, mark size=2pt]
table[x=snr,y=fer]{figures/BLER/v4/DAE32_128_64_xi10.txt};
\addlegendentry{DAE-32-SC} 
    \addplot[color=black,  line width=1.3pt, mark size=2.8pt]
    table[row sep=crcr]{%
    2	0.0224929454291320\\
2.50	0.00453670669501902\\
3	0.000757357807782080\\
3.50	0.000102267269558221\\
4	1.08851674123040e-05\\
4.50	8.87184413330178e-07\\
5	5.35985523713356e-08\\
5.50	2.31423636512729e-09\\
6	6.85484657696988e-11\\
};
    \addlegendentry{Union bound}

    \addplot[color=matlab1_blue, line width=1pt, dashdotted, mark size=2pt]
table[x=snr,y=fer]{figures/BLER/SC_128_99.txt}; 
    \addplot[color=matlab2_red, line width=1pt, mark size=2pt]
table[x=snr,y=fer]{figures/BLER/AE8_128_99.txt};

    \addplot[color=matlab5_green,dotted, mark=x, line width=1pt, mark size=2pt]
table[x=snr,y=fer]{figures/BLER/v4/DAE8_128_99_xi10.txt};

    \addplot[color=matlab4_violet!60,dashed, mark=o, line width=1pt, mark size=2pt]
table[x=snr,y=fer]{figures/BLER/v4/DAE32_128_99_xi10.txt};
    \addplot[color=black,  line width=1.3pt, mark size=2.8pt]
    table[x=snr,y=fer]{figures/BLER/ML_128_99.txt};

\draw[dashed,black,thick]  (axis cs:1.5,0.2) ellipse [    x radius = 20, y radius = 1.6,rotate=110];
    \node [rotate=0] at (axis cs:2.3,0.12) {\footnotesize$(128,64)$};

\draw[dashed,black,thick]  (axis cs:5.1,0.1) ellipse [    x radius = 20, y radius = 1.6,rotate=110];
    \node [rotate=0]at (axis cs:6.2,0.12) {\footnotesize$(128,99)$};
  \end{semilogyaxis}

\end{tikzpicture}%
}
        \caption{BLER performance under DAE for the \gls{rm} codes.}
        \label{fig:DAE_RM}
    \end{figure}
Simulations are for the \gls{awgn} channel with \gls{bpsk} modulation.
The proposed algorithms are denoted as DAE-$M$-SC and PDAE-$M$-SC.
The stopping parameters are optimized for $|\mathcal{B}|=5\times10^5$, $\xi= 10^{-2}$ and a degradation $\epsilon=\nicefrac{\xi}{10}$.
For reference, the upper bound (union bound) of the \gls{ml} performance is also shown. 

    \subsection{Error-correction Performance}

    \autoref{fig:DAE_RM} depicts the error-correction performance under \gls{dae} for the \gls{rm} codes with parameters $(N,K)=(128,64)$ and $(128,99)$.
    Despite designing the parameters $\bm{\sigma}$ with a specific degradation and at the single target BLER of $\xi=10^{-2}$, the degradation remains valid for a wide BLER range.
    By controlling $\xi_D\leq\xi+\nicefrac{\xi}{10}$, the performance of DAE-8-SC and AE-8-SC is very similar.

    \begin{figure}[t]
        \centering
        \resizebox{0.9\columnwidth}{!}{\begin{tikzpicture}
   \pgfplotsset{
     label style = {font=\fontsize{10pt}{7.2}\selectfont},
     tick label style = {font=\fontsize{10pt}{7.2}\selectfont}
   }

  \begin{semilogyaxis}[%
    width=\columnwidth,
    height=0.625\columnwidth,
    xmin=0.5, xmax=6.88,
    xtick={1,1.5,2,...,7.5},
    xlabel={$\text{SNR},\,\mathrm{dB}$},
    xlabel style={yshift=0.4em},
    ymin=4e-5, ymax=0.6,
    ylabel style={yshift=-0.1em},
    ylabel={BLER},
    yminorticks, xmajorgrids,
    ymajorgrids, yminorgrids,
    legend style={at={(0.01,0.01)},anchor=south west},
    legend style={legend columns=1, font=\scriptsize, row sep=-1.2mm},
    legend style={fill=white, fill opacity=1, draw opacity=1,text opacity=1}, 
    legend style={inner xsep=0pt, inner ysep=0pt}, 
    legend cell align={left}, 
    mark size=1.6pt, mark options=solid,
    ]

    \addplot[color=matlab1_blue, line width=1pt, dashdotted]
table[x=snr,y=fer]{figures/BLER/SC_128_60.txt}; 
\addlegendentry{SC}
    \addplot[color=matlab2_red, line width=1pt, mark size=2.1pt]
table[x=snr,y=fer]{figures/BLER/AE8_128_60.txt};
\addlegendentry{AE-8-SC}
    \addplot[color=matlab5_green,dotted, mark=x, line width=1pt, mark size=2pt]
table[x=snr,y=fer]{figures/BLER/v4/DAE8_128_60_xi10.txt};
\addlegendentry{DAE-8-SC} 
\addplot[color=matlab4_violet, line width=1pt, mark size=2pt]
table[x=snr,y=fer]{figures/BLER/AE32_128_60.txt};
\addlegendentry{AE-32-SC}
\addplot[color=matlab4_violet!60,dashed, mark=o, line width=1pt, mark size=2pt]
table[x=snr,y=fer]{figures/BLER/DAE32_128_60.txt};
\addlegendentry{DAE-32-SC}
    \addplot[color=black,  line width=1.3pt, mark size=2pt]
table[x=snr,y=fer]{figures/BLER/ML_128_60.txt};
    \addlegendentry{Union bound}

    \addplot[color=matlab1_blue, line width=1pt, dashdotted]
table[x=snr,y=fer]{figures/BLER/SC_128_98.txt}; 
    \addplot[color=matlab2_red, line width=1pt, mark size=2.1pt]
table[x=snr,y=fer]{figures/BLER/AE8_128_98.txt};

    \addplot[color=matlab5_green,dotted, mark=x, line width=1pt, mark size=2pt]
table[x=snr,y=fer]{figures/BLER/v4/DAE8_128_98_xi10.txt};

    \addplot[color=matlab4_violet, line width=1pt, mark size=2pt]
table[x=es,y=fer]{figures/Results/AE32_128_98.txt};

\addplot[color=matlab4_violet!60,dashed, mark=o, line width=1pt, mark size=2pt]
table[x=snr,y=fer]{figures/BLER/v4/DAE32_128_98_xi10.txt};
    \addplot[color=black,  line width=1.3pt, mark size=2.8pt]
    table[x=snr,y=fer]{figures/BLER/ML_128_98.txt};

\draw[dashed,black,thick]  (axis cs:1.7,0.1) ellipse [    x radius = 20, y radius = 1.7,rotate=110];
    \node [rotate=0]at (axis cs:2.45,0.32) {\footnotesize$(128,60)$};

\draw[dashed,black,thick]  (axis cs:5.1,0.1) ellipse [    x radius = 20, y radius = 1.6,rotate=110];
    \node [rotate=0]at (axis cs:6.,0.3) {\footnotesize$(128,98)$};
  \end{semilogyaxis}

\end{tikzpicture}%
}
        \caption{BLER performance under DAE for the polar codes.}
        \label{fig:DAE_polar}
    \end{figure}
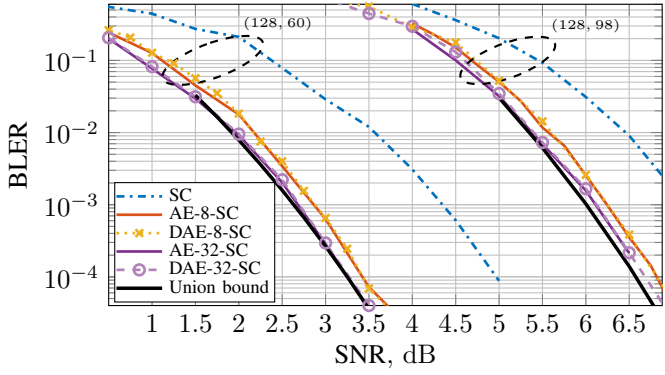

\autoref{fig:DAE_polar} illustrates the error-correction performance of the \gls{dae} for the polar codes with parameters $(128,60)$ and $(128,98)$.
    As for \gls{rm} codes, the parameters $\bm{\sigma}$ are designed solely for $\text{BLER}=10^{-2}$.
    The performance degradation allowed in the construction of $\bm{\sigma}$ dictates the degradation with respect to the  AE-$M$-SC.
    
    \subsection{Complexity Reduction}
The average number $\overline{M}$ of sub-decoders that are used is estimated through simulations.
For $M=\{8,32\}$, \autoref{tab:complexities} summarizes $\overline{M}$ and the complexity reduction factor $\nicefrac{M}{\overline{M}}$ for all investigated codes at $\xi=10^{-3}$. For all codes, $\overline{M}$ is much reduced compared to $M$, showing the effectiveness of our approach for different codes and $M$. Furthermore, for $M=8$, the complexity of DAE approaches that of the FO reference.

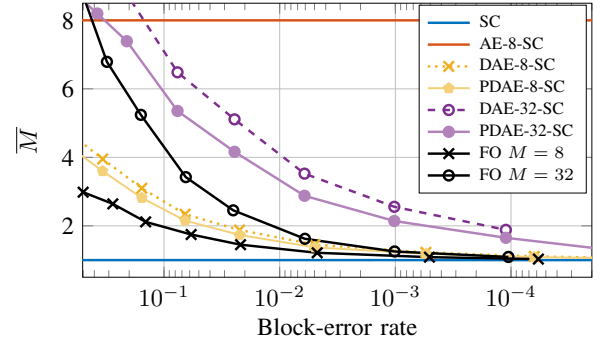
\begin{figure}[t]
        \centering
        \resizebox{0.9\columnwidth}{!}{\usetikzlibrary{spy}
\begin{tikzpicture}[spy using outlines={circle, magnification=2, connect spies}]

  \pgfplotsset{
    label style = {font=\fontsize{10pt}{7.2}\selectfont},
    tick label style = {font=\fontsize{10pt}{7.2}\selectfont}
  }
   \begin{axis}[%
    width=\columnwidth,
    height=0.62\columnwidth,
    xmin=2e-05, xmax=5.05e-01,
    xlabel={Block-error rate},
    xlabel style={yshift=0.4em},
    ymin=0.5, ymax=8.5,
    ytick={2,4,...,16},
    yticklabels={2,4,6,8,10,12,14,16},
    x dir=reverse,
    xmode=log,
    ylabel style={yshift=-1.2em},
    ylabel={$\overline{M}$},
    xlabel style={yshift=-0.2em},
    yminorticks, xmajorgrids,
    ymajorgrids, yminorgrids,
    legend style={at={(1.00,1.00)},anchor=north east},
    legend style={legend columns=1, font=\small, column sep=0.5mm, row sep=-0.5mm, legend cell align=left}, 
    mark size=1.8pt, mark options=solid,
    ] 
  
  \addplot[color=matlab1_blue,  mark=none, line width=1pt] table[row sep=crcr]{%
    1 1\\    
    1e-6 1\\    
    };
  \addlegendentry{SC}
\addplot[color=matlab2_red,  mark=none, line width=1pt] table[row sep=crcr]{%
    1 8\\    
    1e-6 8\\    
    };
  \addlegendentry{AE-8-SC}
  \addplot[color=matlab5_green,dotted,mark=x, line width=1pt,mark size=3pt] table[x=fer,y=averagesubdecoder]{figures/BLER/v4/DAE8_128_64_xi10_Optimal.txt};
  \addlegendentry{DAE-8-SC}

  \addplot[color=matlab5_green!60,dotted,mark=pentagon*,solid, line width=0.8pt,mark size=2pt] table[x=fer,y=averagesubdecoder]{figures/BLER/v4/DPAE8_128_64_xi10_Optimal.txt};
    \addlegendentry{PDAE-8-SC}

  \addplot[color=matlab4_violet,dashed, mark=o, line width=1pt, mark size=2pt] table[x=fer,y=avgADEC]{figures/Results/DAE32_128_64_new.txt};
    \addlegendentry{DAE-32-SC}

  \addplot[color=matlab4_violet!60,solid, mark=*, line width=1pt, mark size=2pt] table[x=fer,y=avgADEC]{figures/Results/DPAE32_128_64_new.txt};
    \addlegendentry{PDAE-32-SC}

  \addplot[color=black,solid, line width=1pt,mark=x,mark size=3pt] table[x=fer,y=FO]{figures/Results/AE8_128_64.txt};
    \addlegendentry{FO $M=8$}

      \addplot[color=black,solid, mark=o,line width=1pt,mark size=2pt] table[x=fer,y=FO]{figures/Results/AE32_128_64.txt};
    
\addlegendentry{FO $M=32$}

     \coordinate (spypoint2) at (axis cs:0.0001,1.25); 
  \coordinate (magnifyglass2) at (axis cs:0.011,6); 

     \coordinate (spypoint) at (axis cs:0.001,1.25); 
  \coordinate (magnifyglass) at (axis cs:0.11,6); 
\end{axis}    
\end{tikzpicture}}
        \caption{Average number $\overline{M}$ of sub-decoders executed in \gls{dae} and \gls{pdae} for the $(128,64)$ \gls{rm} code.}
        \label{fig:DAE_energy}
    \end{figure}

\autoref{fig:DAE_energy} illustrates \( \overline{M} \) for $M=8$ and $M=32$ as a function of the \gls{bler} for the RM code \( (128,64) \).
The complexity of the proposed \gls{dae} and \gls{pdae} algorithms decreases with the \gls{bler}.
Despite designing the parameters $\bm{\sigma}$ at a single \gls{bler} reference, we see that the early stopping provides a large complexity reduction over a wide range of \glspl{bler}.
The \gls{pdae} approach provides a greater gain for $M=32$ which can be explained by the larger ensemble and, due to the \gls{bler} improvements, the lower SNR.

\begin{table}[t]
	\centering
	\caption{Complexity at $\xi=10^{-3}$ for $M=\{8,32\}$} 
	\label{tab:complexities}
	{\begin{tabular}{|c|c|c|c|c|c|}
		\cline{1-5}
        \rule{0pt}{10pt}Decoding & \multicolumn{4}{c|}{Code $(N,K)$} &\multicolumn{1}{c}{}\\
		\cline{2-5}
        \rule{0pt}{10pt}complexity & $(128,60)$ & $(128,64)$ & $(128,98)$ & $(128,99)$ &\multicolumn{1}{c}{}\\
		\hline
        \multicolumn{1}{|c|}{$\Mfo$} & 1.069 & \change{1.129} &\change{1.034}  & \rule{0pt}{10pt}\change{1.047}&\multirow{5}{*}{\rotatebox{90}{M=8}}\\
        \cline{1-5}
         \rule{0pt}{10pt} $\overline{M}$ DAE & 1.35 & 1.28& 1.145 &1.184 & \\
          $\nicefrac{M}{\overline{M}}$ DAE & 5.9$\times$  & 6.3$\times$& 7.0$\times$ &6.8$\times$ &\\
         \cline{1-5}
         \rule{0pt}{10pt}$\overline{M}$ PDAE &1.28 & 1.23& 1.101&1.141 &\\
		$\nicefrac{M}{\overline{M}}$ PDAE & 6.3$\times$ &6.5$\times$ & 7.3$\times$&7.0$\times$ &\\
		  \hline
        \hline
		  \multicolumn{1}{|c|}{$\Mfo$} & 1.115  & \rule{0pt}{10pt}\change{1.254}& 1.058 & \change{1.093}&\multirow{5}{*}{\rotatebox{90}{M=32}}\\
		  \cline{1-5}
         \rule{0pt}{10pt}$\overline{M}$ DAE & 2.41& 2.557 & 1.843&1.66 & \\
         $\nicefrac{M}{\overline{M}}$ DAE & 13.2$\times$ & 12.5$\times$ & 17.4$\times$& 19.3$\times$& \\
         \cline{1-5}
         \rule{0pt}{10pt}$\overline{M}$ PDAE &2.08 & 2.142& 1.552 &1.46 &\\
         $\nicefrac{M}{\overline{M}}$ PDAE & 15.4$\times$& 14.9$\times$& 20.6$\times$ & 21.9$\times$ &\\
		\hline
	\end{tabular}}
\end{table}

    \section{Conclusions}\label{sec:concl}
In this letter, we proposed a sequential dynamic automorphism ensemble decoding algorithm.
To this end, the redundancy inherent to the original \gls{ae} decoder was examined, and the correlation between the decoding outcome of each sub-decoder and the magnitude of the associated path metric was investigated.
An optimization algorithm for determining early-termination parameters was introduced and the resulting parameters were shown to be effective over a wide range of \glspl{bler}.
Verified with polar and \gls{rm} codes, the proposed approach preserves the error-correction performance while reducing the decoding complexity by at least $6\times$ compared to the original \gls{ae} decoder for \glspl{bler} below $10^{-3}$.

\bibliographystyle{IEEEtran}
\bibliography{IEEEabrv,ConfAbrv,references}
\end{document}